ORIGINAL ARTICLE

# Combating Financial Crimes With Unsupervised Learning Techniques: Clustering and Dimensionality Reduction for Anti-Money Laundering

Ahmed Nagy Bakry [a],*, Almohammady Sobhy Alsharkawy [a], Mohamed Sayed Farag [a,b], Kamal Raslan Mohamed Raslan [a]

[a] Department of Mathematics, Faculty of Science, Al-Azhar University, Cairo, Nasr City, Egypt
[b] Obour Height Institute for Informatics, Cairo, Egypt

## Abstract

Anti-Money Laundering (AML) is a crucial task in ensuring the integrity of financial systems. One key challenge in AML is identifying high-risk groups based on their behavior. Unsupervised learning, particularly clustering, is a promising solution for this task. However, the use of hundreds of features to describe behavior results in a high-dimensional dataset that negatively impacts clustering performance. In this paper, we investigate the effectiveness of combining clustering method agglomerative hierarchical clustering with four dimensionality reduction techniques -Independent Component Analysis (ICA), and Kernel Principal Component Analysis (KPCA), Singular Value Decomposition (SVD), Locality Preserving Projections (LPP)- to overcome the issue of high-dimensionality in AML data and improve clustering results. This study aims to provide insights into the most effective way of reducing the dimensionality of AML data and enhance the accuracy of clustering-based AML systems. The experimental results demonstrate that KPCA outperforms other dimension reduction techniques when combined with agglomerative hierarchical clustering. This superiority is observed in the majority of situations, as confirmed by three distinct validation indices.

*Keywords:* Anti-money laundering, Clustering, Dimension reduction, Cluster verification

## 1. Introduction

Anti-money laundering (AML) is a crucial task in the financial sector as it helps prevent the illegal flow of funds that can finance terrorism, drug trafficking, and other crimes. One traditional approach to AML is the use of rule-based models, which rely on predefined rules and thresholds to identify suspicious transactions. Rule-based models in AML use predefined criteria to flag transactions that may be indicative of money laundering activities. For example, a rule-based model may flag a transaction if it exceeds a certain dollar amount or if it originates from a country with a high risk of money laundering. The rules used by rule-based models are often determined by regulatory bodies and are based on historical data and expert knowledge.

While rule-based models can be effective in some contexts, they also have limitations. For example, the rules used by these models may not be able to keep pace with the constantly evolving tactics used by money launderers. Additionally, rule-based models can result in a high number of false positives, as they may flag legitimate transactions as suspicious [1]. This can result in increased operational costs, as financial institutions must investigate each false positive to determine if it is indeed a case of money laundering.

Machine learning models have the advantage of being able to learn from data and adapt to changing patterns of money laundering activities, potentially improving the efficiency and effectiveness of AML efforts. For example, a machine learning model may be able to identify subtle patterns in transaction data that would be difficult for a rule-based model to detect. Additionally, machine learning models can be trained on large amounts of historical data, allowing them to identify new and emerging money laundering techniques.

However, the development and deployment of machine learning models for AML also come with their own set of challenges. One of the main challenges is the need for large amounts of labeled data to train the models. This can be a challenge in the context of AML, as the labeling of transaction data can be time-consuming and resource-intensive. Additionally, machine learning models have the potential to introduce bias into the AML process, particularly if the training data is biased in some way. To minimize the risk of bias, it is important to carefully consider the sources of data used to train machine learning models for AML and to monitor the performance of these models over time [2].

Unsupervised learning techniques, such as clustering, can be used to identify patterns and relationships in the data that may not be apparent through traditional rule-based approaches. Clustering algorithms can group together transactions that are similar to each other based on a set of features, such as the amount of money being transferred, the location of the sender and receiver, and the time of the transaction. This can help to identify groups of transactions that may be indicative of money laundering activities, even if these transactions do not fit the criteria of a traditional rule-based model.

To further improve the efficiency of clustering algorithms in the AML context, Dimension Reduction Techniques (DRTs) can be applied to reduce the dimensionality of the data. This can help to remove noise and redundancy in the data, making it easier for the clustering algorithms to identify meaningful patterns. Some of the most commonly used DRTs in the context of AML include ICA, KPCA, SVD and LPP. The contributions of this paper are summarized as follows:

- Investigate the possibilities of unsupervised machine learning techniques, specifically clustering, in order to tackle the challenge of AML.
- Analyze the impact of DRTs on the performance of agglomerative hierarchical clustering.
- Utilize the solution in the context of real-world data and provide a novel approach for profiling said data.
- Discuss validation indices and their potential influence while utilizing DRT.

The organization of this paper is presented as follows: in Section 2, we present overviews of related works. Then, Section 3 is devoted to the presentation of the proposed methodology. Section 4 describes our experimental evaluation approach and the different parameters used, along with a discussion of the results obtained. Finally, Section 5 concludes this paper.

## 2. Related work

In recent years, the use of machine learning techniques for AML purposes has gained growing interest. Unsupervised learning, in particular, has emerged as a promising approach as it can identify patterns and relationships in the data without labeled examples. Clustering algorithms, a common type of unsupervised learning, have been applied in the AML context.

Several studies have investigated the use of clustering algorithms for AML, including one [3] which aimed to evaluate the effectiveness of these techniques in detecting anomalous payments. The authors used transactional information from a payment system in El Salvador, preprocessed the network features (such as degree and strength) using principal component analysis to reduce dimensionality, and applied the k-means and DBSCAN algorithms to identify clusters of anomalous payments. The results were further validated using the random forest algorithm. The study analyzed network and amount features, including amount, beneficiary counterparty degree, origin participant degree, and beneficiary participant strength, but only focused on wire type payments and did not analyze profiling features over a long-term period. Another study [4] aimed to enhance the AML system by optimizing the rule-based model, resulting in a decrease of 18 % in false positives while maintaining 98 % of true positives. However, this study did not take into account different customer behaviors, potentially oversimplifying the AML process and missing important indicators of money laundering.

In another study [5], the authors present a unique approach to assessing the EU legal framework in terms of money laundering prevention by using clustering to group EU member states based on their money laundering measuring indices. This approach provides a more comprehensive understanding of the money laundering situation in each country and how it relates to other factors such as the number of suspicious transaction reports, the Basel AML Index, and the Corruption Perception Index. The authors highlight the negative impact of money laundering on economic development and provide practical recommendations for policy-makers and anti-money laundering institutions. However, the study has limitations such as being based on data from a specific time period, only focusing on EU member states, and not being generalizable to other time periods.

Another paper [6] presents Amaretto, an active learning system for anomaly detection applied to transaction monitoring for money laundering detection in capital markets. The system uses an unsupervised model to detect both known and unknown anomalous patterns, employs four strategies to optimally sample data for review by a subject-matter expert, and feeds the reviewed data into a supervised learning model to improve performance. The experiments conducted on a synthetic dataset that resembles genuine and potential money laundering patterns show that Amaretto achieved state-of-the-art performance in a short time frame with minimal manual input. However, the use of synthetic data instead of real data and the lack of an intuitive explanation for the anomaly score used to rank high-level vectors are limitations of the research that may impact its generalizability.

The authors in [7] explore the use of client profiling for an AML system. This paper uses account movement databases to create client profiles, clusters, and rules that can identify suspicious transactions. The study found that using a customized definition of client profiles and a larger, longer-term dataset improved results in terms of cluster evaluation metrics and generated rules. The results were verified by a financial institution specialist. However, the study lacks a description of the feature selection process. The literature review in [8] addresses the challenges in the field of AML and overviews various machine learning techniques used to overcome them. The authors identify two main challenges: class imbalance and lack of publicly available datasets, and propose using synthetic data as a solution. The authors also acknowledge that financial institutions have access to high-dimensional and unlabeled data, requiring the use of dimension reduction and semi-supervised learning. The review covers other potential research areas such as data visualization, deep learning, and interpretable and fair machine learning, but does not provide in-depth analysis of specific challenges and solutions for each area.

The paper [9] employs k-means clustering to analyze customer behavior in anti-money laundering systems using customer transaction data from a one-year period. The study used 11 features to create customer profiles and perform the clustering. A limitation of this study is the limited number of features used, which may not have captured all relevant information about customer behavior and thus impacted the accuracy of the clustering analysis. The use of only one year of transaction data also restricts the scope of the study and the insights it can provide into customer behavior over time. In another study [10], investigates the effectiveness of two DRTs, Linear Discriminant Analysis (LDA) and Principal Component Analysis (PCA), when applied to a large dataset. The study evaluates the performance of these techniques using four machine learning models on a Cardiotocography (CTG) dataset. The results showed that PCA outperformed LDA in all measures, without affecting the results of the machine learning models used. The authors also found that using high-dimensional PCA with machine learning algorithms led to better results compared to using machine learning with all dimensions. However, using low-dimensional PCA resulted in lower performance compared to using machine learning with all dimensions.

This paper [11] provides an overview and comparative study of dimensionality reduction techniques for high-dimensional data. The study examines both linear and non-linear dimensionality reduction techniques and focuses on the selection of suitable techniques for diverse types of data, including text, numeric, signals, etc. It also investigates open issues associated with dimensionality reduction techniques in different applications and explores solutions to high dimensional data issues using appropriate dimensionality reduction techniques. The examination is performed on an ElectroCardioGram (ECG) signal for heartbeat data derived from the PhysioNet MIT-BIH Arrhythmia database, which consists of 188 dimensions and 1 million observations. The data set is used to evaluate linear dimension reduction techniques, such as Principal Component Analysis (PCA), Singular Value Decomposition (SVD), Latent Semantic Analysis (LSA) for text, Locality Preserving Projections (LPP), Independent Component Analysis (ICA), and Projection Pursuit (PP), and non-linear dimensionality reduction techniques, including Kernel Principal Component Analysis (KPCA), Multidimensional Scaling (MDS), Isomap, Locally Linear Embedding (LLE), Self-Organizing Map (SOM), Learning Vector Quantization (LVQ), and T-Stochastic Neighbor Embedding (TSNE). In conclusion, the related work on anti-money laundering and DRTs has shown the importance of utilizing large and real data sets, as well as the various methods available for reducing high-dimensional data.

The previous studies have demonstrated that different techniques may perform better or worse, depending on the specific data set and application. This highlights the importance of considering a range of methods in order to find the most effective solution for a particular problem. The examination of different clustering techniques, such as agglomerative hierarchical clustering and linear and nonlinear dimensionality reduction methods, provides a valuable foundation for our ongoing research in the field of anti-money laundering and the use of machine learning for this purpose. We will be applying these findings to our own study and evaluating the effectiveness of various DRTs on agglomerative hierarchical clustering as a clustering technique in order to enhance the performance of our anti-money laundering system.

## 3. Methodology

In this research, we aim to profile the historical transactions of customers over the last year and use this information to cluster their behavior. Our profile contains information about all transaction types such as cash, wire, and check, along with the averages, minimums, and maximums of each type. This high-dimensional dataset presents challenges for clustering, so we apply DRTs to reduce the dimensionality of the data. In addition, various clustering verification techniques are employed to determine the most suitable combination of DRT and validation method for a given clustering technique.

The agglomerative hierarchical clustering will be applied on the dataset to determine the best DRT for clustering customer behavior based on the cluster vitrification scores. This involves analyzing the results of the DRTs and comparing the performance of each method to find the best solution. Our goal is to find the optimal framework for profiling customer behavior in anti-money laundering systems.

### 3.1. Dimension reduction techniques (DRT)

In this subsection, we will explore and compare several DRTs for profiling historical transactions of customers in an anti-money laundering system. The techniques include ICA, KPCA, SVD and LPP. These methods are used to reduce the high-dimensional feature space of the customer transaction data, which will then be used as input for clustering methods. The objective is to identify the best DRTs that can accurately capture the underlying structure of the customer transaction data, while minimizing the loss of information.

#### 3.1.1. Independent Component Analysis (ICA)

The ICA [12,13] is a statistical technique for separating independent sources from observed mixtures of signals. It is used to decompose a multivariate signal into independent non-Gaussian components. In contrast to principal component analysis (PCA) or linear discriminant analysis (LDA), ICA does not assume that the data components are Gaussian or linear, but rather models them as independent, non-Gaussian and non-linear. The standard equation for ICA is given by:

$$\mathbf{x} = \mathbf{As} \quad (1)$$

Where $\mathbf{x}$ is the observed signal, $\mathbf{s}$ is the underlying independent component, and $\mathbf{A}$ is the mixing matrix. The goal of ICA is to find the demixing matrix $\mathbf{W}$ that separates the independent components, such that:

$$\mathbf{s} = \mathbf{Wx} \quad (2)$$

In mathematical terms, the goal of ICA is to maximize the non-Gaussianity of the transformed signals $\mathbf{s}$. This is done by maximizing the entropy or non-Gaussianity of the transformed signals, which can be achieved by minimizing the kurtosis or maximizing the negentropy of the transformed signals.

#### 3.1.2. Kernel Principal Component Analysis (KPCA)

The KPCA [14,15] is a non-linear dimensionality reduction technique that maps the input data into a lower-dimensional space by transforming the input data into a high-dimensional feature space, and then finding the principal components in the transformed feature space. KPCA with radial basis function (RBF) kernel is one popular implementation of KPCA.

The basic idea behind KPCA is to map the original data into a high-dimensional space where the non-linear structure of the data can be captured and represented in a linear form. The RBF kernel is used to calculate the inner product between any two data points in the feature space. The principal components are then obtained by performing a singular value decomposition (SVD) on the covariance matrix of the data in the feature space. The equation for RBF kernel used in KPCA can be represented as follows:

$$k(x,y) = \exp\left(-\frac{|x-y|^2}{2\sigma^2}\right) \quad (3)$$

Where $k(x,y)$ is the RBF kernel between the two data points $x$ and $y$, $|x-y|^2$ is the squared Euclidean distance between $x$ and $y$, and $\sigma$ is a free parameter that determines the width of the Gaussian function. The choice of $\sigma$ affects the non-linear mapping of the data.

### 3.1.3. Singular Value Decomposition (SVD)

The first dimension reduction technique that we applied in this research is SVD [16]. SVD is a linear algebraic method that decomposes a high-dimensional matrix into three matrices: the left singular vectors, the singular values, and the right singular vectors. The decomposition process results in a lower-dimensional representation of the original data, preserving important information while reducing the number of dimensions. SVD is a powerful technique that can be used for various purposes, including dimensionality reduction, data compression, and data denoising. The standard equation for SVD is as follows:

$$A = U\Sigma V^T \quad (4)$$

Where $A$ is an $m \times n$ matrix, $U$ is an $m \times m$ orthogonal matrix, $\Sigma$ is an $m \times n$ diagonal matrix containing the singular values, and $V$ is an $n \times n$ orthogonal matrix. By selecting the top $k$ singular values, SVD reduces the dimensions of the original matrix $A$ to $k$.

### 3.1.4. Locality Preserving Projections (LPP)

The LPP [17] is a linear dimensionality reduction technique that is commonly used for preserving the local structure of the high-dimensional data. The basic idea behind LPP is to embed the high-dimensional data into a low-dimensional space while preserving the local neighborhood relationships between data points. This is achieved by minimizing the reconstruction error between the high-dimensional data and its low-dimensional representation. The objective function of LPP can be represented as follows:

$$\min_{\mathbf{W}} tr(\mathbf{W}^T \mathbf{L} \mathbf{W}) \quad (5)$$

Where $\mathbf{W}$ is the projection matrix and $\mathbf{L}$ is the Laplacian matrix of the data graph. The Laplacian matrix can be calculated by finding the affinity matrix between data points and constructing a diagonal matrix based on the sum of the affinity values. LPP can effectively capture the non-linear structures in the high-dimensional data, and it has been widely used in various applications, such as image recognition, text classification, and face recognition.

### 3.2. Clustering method

Clustering is a machine learning technique that groups similar data points together based on their feature similarity. It is an unsupervised learning approach, as the data points are not labeled beforehand. There are several clustering methods that can be applied to data, including K-means, Agglomerative Hierarchical Clustering, and Density-Based Spatial Clustering of Applications with Noise (DBSCAN). In this research, Agglomerative Hierarchical Clustering is used to cluster the reduced feature set obtained from the dimensionality reduction techniques.

The Agglomerative Hierarchical Clustering (AHC) is a bottom-up approach to clustering [18], where individual data points are merged together to form clusters. The basic idea is to start with each data point as its own cluster, and then merge the closest pair of clusters into a single cluster until a stopping criterion is met. The final result is a tree-like structure, known as a dendrogram, that represents the hierarchical structure of the clusters. AHC can be described as an algorithm with the following steps:

1) Start by treating each data point as its own cluster.
2) Compute the distance matrix between all pairs of clusters.
3) Merge the two closest clusters into a single cluster.
4) Repeat step 2 and 3 until a stopping criterion is met (e.g., a maximum number of clusters or a maximum height of the dendrogram).
5) Extract the clusters from the dendrogram based on a suitable distance threshold.

### 3.3. Clustering verification indices

After performing clustering on the dataset, it is important to evaluate the quality of the clustering. Cluster validation or verification is the process of evaluating the goodness of the clusters obtained from clustering algorithms. This step is essential to choose the best clustering algorithm and the optimal number of clusters. In this subsection, we will discuss some popular clustering indices that are used to evaluate the quality of the clustering.

#### 3.3.1. Silhouette score

[19] is a popular clustering validation index that measures how well each data point is assigned to its assigned cluster, and the quality of the clustering results as a whole. It calculates the average distance

between each data point to its own cluster, compared to the distance between the same data point and the nearest neighboring cluster. The silhouette score ranges from −1 to 1, where a higher score indicates that the data point is well-matched to its own cluster, and poorly matched to neighboring clusters. The silhouette score for each data point $i$ is calculated as follows:

$$s(i) = \frac{b(i) - a(i)}{max(a(i), b(i))} \quad (6)$$

Where $a(i)$ is the average distance between data point $i$ and all other data points in the same cluster, and $b(i)$ is the average distance between data point $i$ and all other data points in the nearest neighboring cluster. The overall silhouette score is the average of all silhouette scores.

The value of the silhouette score ranges from −1 to 1, where a value closer to 1 indicates that the clustering is better. A negative score indicates that the data point may have been assigned to the wrong cluster. The silhouette score can be used to compare the quality of different clustering results and choose the optimal number of clusters.

### 3.3.2. Calinski-Harabasz Score

[20] is another clustering index used to evaluate the quality of clusters. It is also known as the Variance Ratio Criterion. The Calinski-Harabasz index measures the ratio of the between-cluster dispersion and the within-cluster dispersion. It is defined as the ratio of the sum of squares of the distances between the cluster centroids and the total sum of squares. A higher Calinski-Harabasz index indicates better-defined clusters. The Calinski-Harabasz score can be calculated using the following formula:

$$CH(K) = \frac{tr(B_K)}{tr(W_K)} \times \frac{N - K}{K - 1} \quad (7)$$

Where K is the number of clusters, N is the total number of samples, $tr(B_K)$ is the trace of the between-cluster scatter matrix, and $tr(W_K)$ is the trace of the within-cluster scatter matrix. The within-cluster scatter matrix is defined as:

$$W_K = \sum_{q=1}^{K} \sum_{x \in C_q} (x - \mu_q)(x - \mu_q)^T \quad (8)$$

Where $C_q$ is the set of data points in the qth cluster, $\mu_q$ is the mean of the data points in $C_q$, and T denotes the transpose. The between-cluster scatter matrix is defined as:

$$B_K = \sum_{q=1}^{K} N_q (\mu_q - \mu)(\mu_q - \mu)^T \quad (9)$$

Where $N_q$ is the number of data points in $C_q$, $\mu$ is the overall mean of the data, and $\mu_q$ is the mean of the data points in $C_q$. The Calinski-Harabasz index returns a score that can be used to compare different clustering results. A higher score indicates better-defined clusters.

### 3.3.3. Davies-Bouldin Score

[21] is another clustering validation index that measures the ratio of the within-cluster scatter to the between-cluster separation. It is calculated as the average similarity between each cluster and its most similar cluster, where similarity is defined as the ratio of the sum of within-cluster distances to the between-cluster distance. The index ranges from 0 to infinity, with lower values indicating better clustering. The formula for calculating the Davies-Bouldin index for a set of clusters $C = C_1, C_2, \ldots, C_k$ is:

$$DB = \frac{1}{k} \sum_{i=1}^{k} \max_{j \neq i} \left( \frac{\sigma_i + \sigma_j}{d(c_i, c_j)} \right) \quad (10)$$

Where $\sigma_i$ is the average distance between each point in cluster $C_i$ and the centroid $c_i$ of that cluster, and $d(c_i, c_j)$ is the distance between the centroids $c_i$ and $c_j$ of clusters $C_i$ and $C_j$. The Davies-Bouldin index is minimized when each cluster is compact and well-separated from other clusters.

## 4. Experiments and results

This section focuses on the validation and evaluation of the selected four-dimensionality reduction techniques along with hierarchical clustering in the context of AML using customer transaction data. The DRTs are evaluated with different numbers of components and various segments in terms of three clustering verification methods (Silhouette Score, Calinski-Harabasz Score, and Davies-Bouldin Score). The experiments were conducted on Google Colaboratory, which provides a free notebook environment with support for running machine learning experiments. The details of these experiments are as follows:

### 4.1. Experimental dataset

Our experiments are conducted using a dataset obtained from the banking industry. Due to confidentiality considerations, we are unable to disclose the specific identity of the bank or furnish exact details. Nevertheless, we are able to offer approximations in order to describe the data where it is possible for us to do so.

### 4.1.1. Data collection

Transaction data from various customers were collected. The data was organized into yearly profiles for each customer, with a total of 4099 companies included in the dataset. Each company profile was characterized by 80 features, resulting in a dataset with 4099 rows and 80 columns. A comprehensive summary of the data may be found in Table 1.

### 4.1.2. Yearly customer profiles

The yearly customer profile is a critical component of our experimental design Fig. 1, aimed at capturing and summarizing customer transaction behavior over a one-year period. Transactions are categorized into different types, including but not limited to wires, cash, and checks. For each transaction type, both credit and debit transactions are considered. To construct a comprehensive customer profile, we extract five key measures for each transaction type:

1) **Minimum (Min)**: The smallest transaction amount within a specific category during the year.
2) **Maximum (Max)**: The largest transaction amount within a specific category throughout the year.
3) **Average (Avg)**: The mean transaction amount for a particular transaction type.
4) **Count (Cnt)**: The total number of transactions in a given category over the year.
5) **Sum (Sum)**: The cumulative sum of transaction amounts for a specific transaction type during the year.

By capturing these five measures for each transaction type, we create a detailed profile that characterizes customer behavior, enabling us to explore and analyze patterns, anomalies, and relationships in the transaction data. This customer profile forms the basis for subsequent feature reduction and clustering experiments, providing valuable insights into customer transaction patterns that are essential for AML efforts. This experimental design outlines the systematic approach to analyzing customer transaction data, reducing feature dimensions, and applying various clustering techniques with different configurations to evaluate their performance.

### 4.2. Experimental configurations

In this phase, we outline the experimental design Fig. 2 conducted to investigate the performance of different DRTs fed into AHC in the context of AML using customer transaction data. The experiments aim to determine the optimal combinations of DRTs, and the number of retained features, as well as the number of clusters, to achieve effective and accurate detection of suspicious transactions. The experimental design encompasses the following key stages: data collection, calculation of yearly customer profiles, application of feature reduction methods, tuning of feature reduction parameters, and adjustment of clustering parameters. Each stage is carefully designed to explore various configurations and evaluate their impact on the quality of the clustering results.

1) **Data Collection:** Transaction data from diverse customers are collected.
2) **Yearly Customer Profiles:** Yearly customer profiles are computed based on these transactions. These profiles encompass various statistics and characteristics of customer transactions over the year.
3) **DRTs:** Several dimension reduction methods are applied to the customer profiles, including ICA, KPCA, SVD and LPP.
4) **Tuning DRTs:** Each DRT is fine-tuned with varying numbers of retained features. Different numbers of features, such as 2, 10, 20, and 40, are explored to assess their impact on subsequent clustering results.
5) **Tuning Clustering Parameters:** For AHC method, various numbers of clusters are examined, such as 3, 5, 7, and 9. The objective is to investigate how the number of clusters influences the quality of clustering results.

The aim is to determine the best appropriate feature reduction approaches and verification procedures for clustering customer profiles for AML purposes, taking into account various scenarios and settings. The outcomes of these trials will yield significant knowledge regarding the development of an efficient AML system capable of effectively identifying and thwarting money laundering endeavours. The subsequent sections provide a comprehensive overview of the experimental design, encompassing the methodology, parameters, and configurations investigated in the study and Table. The experimental arrangement and settings are summarised in Table 2.

Table 1. Data summary.

| Description | Count |
| --- | --- |
| Count of Rows (Companies) | 4099 |
| Count of Features | 80 |
| Count of transaction types | 20 |
| Count of measures for each transaction type | 5 |

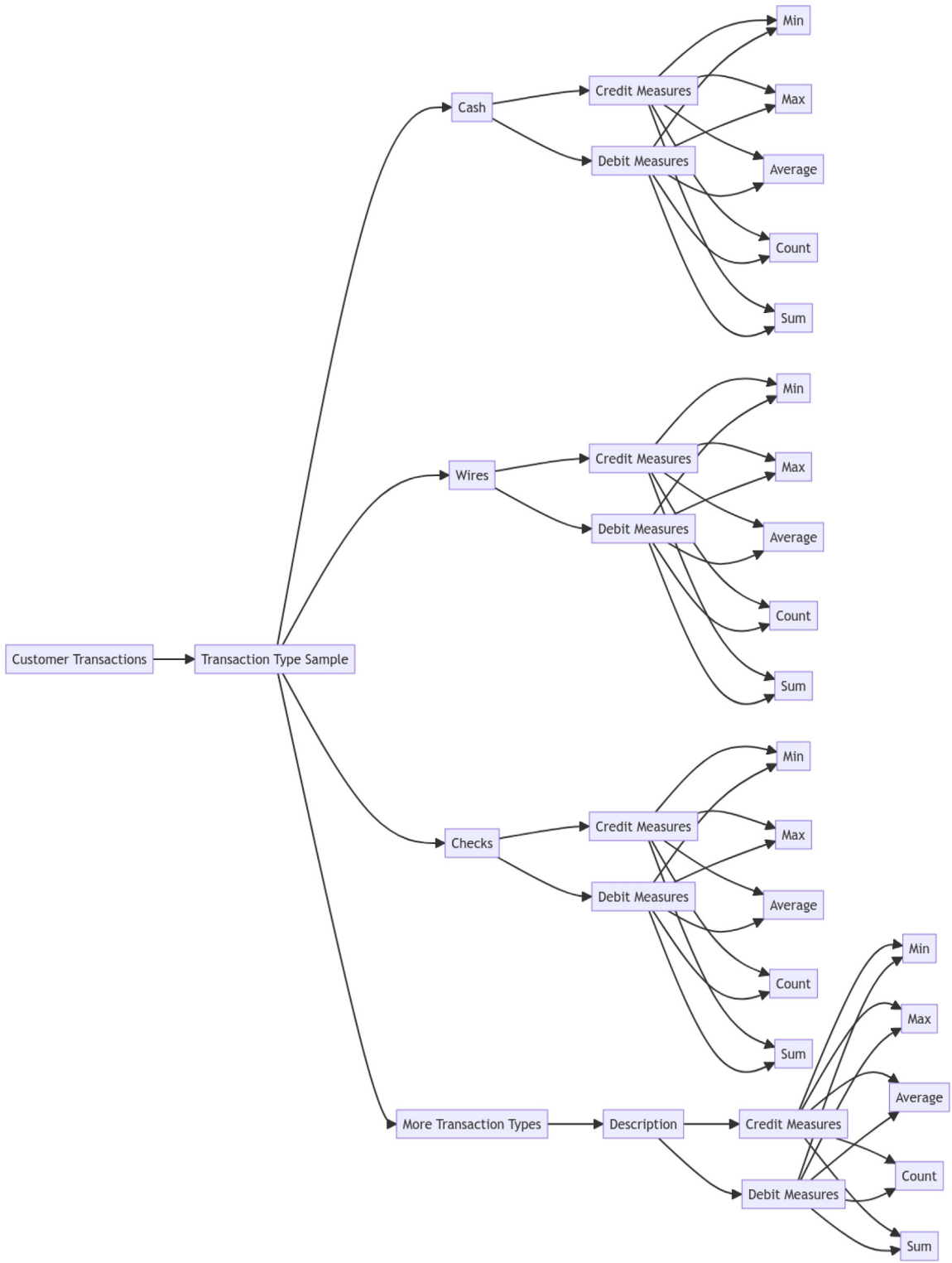

Fig. 1. Customer profile structure.

4.3. Results and discussion

In this section, we shall engage in a comprehensive analysis and discussion of the outcomes and experiments conducted.

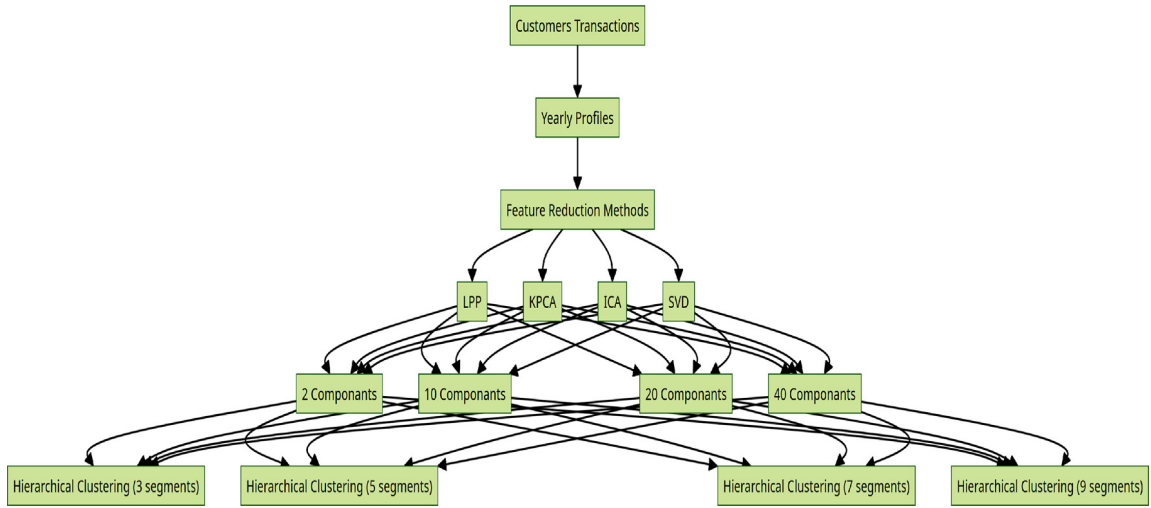

Fig. 2. Experiments design.

### 4.3.1. Hierarchical clustering with three segments

We first evaluated the performance of different dimensionality reduction techniques, specifically ICA, KPCA, SVD, and LPP, with varying numbers of components (C) for a 3-segment clustering task. The following metrics were used for evaluation: Silhouette Score, Davies-Bouldin Index, and Calinski-Harabasz Index. Here are the results:

Table 3 presents the results of applying various DRTs to a dataset intended for 3-segment clustering. The evaluation metrics include the Silhouette Score, Calinski-Harabasz Index, and Davies-Bouldin Index, which collectively assess the quality of clustering outcomes.

Starting with the Silhouette Score, it measures the similarity of data points within the same cluster compared to other clusters. Higher Silhouette Scores indicate better-defined and well-separated clusters. Notably, the KPCA with two components and three Segments configuration achieves a nearly perfect Silhouette Score of 0.9992, signifying excellent cluster separation. Conversely, the SVD and LPP-based configurations exhibit lower Silhouette Scores, indicating less distinct clusters.

The Calinski-Harabasz Index assesses the ratio of between-cluster variance to within-cluster variance,

Table 2. Experimental setup.

| Clustering Techniques | Hierarchical clustering |
| --- | --- |
| DRT's | ICA |
| | KPCA |
| | LPP |
| | SVD |
| Numbers of Components | 2, 10, 20, and 40 |
| Numbers of Clusters (Segments) | 3, 5, 7, and 9 |
| Clustering Verification indices | Sil. score |
| | CH. score |
| | DB. score |

Table 3. Results for clustering verification indices - Configuration 3-segments.

| DRT | # of components | Sil. score | CH. score | DB. Score |
| --- | --- | --- | --- | --- |
| ICA | | 0.8093 | 1183.2046 | 0.4224 |
| KPCA | 2 components | **0.9992** | **12500.48** | **0.0009** |
| SVD | | 0.7898 | 482.9016 | 0.3218 |
| LPP | | 0.7679 | 1268.8426 | 0.0 |
| ICA | | 0.8417 | 43.7997 | 0.0975 |
| KPCA | 10 components | **0.98** | 890.7343 | **0.01** |
| SVD | | 0.8307 | 39.5812 | 0.1107 |
| LPP | | 0.7577 | **1944.10** | 0.0 |
| ICA | | **0.91** | 148.63 | 0.05 |
| KPCA | | 0.83 | 39.58 | 0.11 |
| SVD | 20 components | 0.83 | 39.58 | 0.11 |
| LPP | | 0.76 | 267.05 | 0.0 |
| ICA | | 0.83 | 56.75 | 0.52 |
| KPCA | 40 components | **0.98** | **706.34** | **0.01** |
| SVD | | 0.83 | 39.58 | 0.11 |
| LPP | | 0.76 | 267.05 | 0.00 |
| Without DRT | 80 components | 0.8307 | 39.5812 | 0.1107 |

providing insights into cluster compactness and separation. A higher Calinski-Harabasz Index suggests more compact and well-separated clusters. In this context, KPCA with two components again stands out with the highest score of 12500.4844, emphasizing its effectiveness in forming compact and separated clusters. The SVD and LPP-based configurations display comparatively lower Calinski-Harabasz Index values.

Lastly, the Davies-Bouldin Index measures the average similarity between each cluster and its most similar cluster, with lower values indicating better separation. Here, KPCA with two components and three segments achieves an exceptionally low Davies-Bouldin Index of 0.0009, reinforcing the presence of highly distinct clusters. However, SVD yield higher Davies-Bouldin Index values, suggesting less effective cluster separation.

The DBS (Davies-Bouldin Index) yielding a value of 0.0 in the context of Locality Preserving Projections (LPP) for dimensionality reduction can be attributed to the extreme compactness and isolation of clusters in the reduced feature space. LPP aims to preserve local neighborhood relationships when reducing high-dimensional data to a lower dimension. However, extreme dimensionality reduction can lead to highly distinct and well-separated clusters. In such cases, clusters have minimal overlap, making them appear entirely dissimilar to the DBS index, which computes average inter-cluster similarity.

While a DBS score of 0.0 might suggest effective clustering, it is essential to consider the broader context and specific analysis goals. Extreme dimensionality reduction may align with certain objectives but may not reflect the desired cluster properties in other scenarios. Therefore, interpreting a DBS score of 0.0 should be done cautiously, considering the practical implications and domain-specific knowledge when assessing clustering quality.

*4.3.2. Hierarchical clustering with five segments*

From Table 4, we observe that the best Silhouette Scores are achieved by the methods ICA with twenty components (0.87), ICA with 10 components (0.83), and KPCA with ten components (0.99). These configurations exhibit well-separated clusters, as indicated by their high Silhouette Scores.

Conversely, configurations such as LPP with two components and LPP with ten components have lower Silhouette Scores (0.72 and 0.74, respectively), suggesting less distinct clusters. Notably, KPCA with two components stands out with an exceptionally high Silhouette Score of 0.9999, indicating near-perfect clustering for this specific configuration. From Table 4, it is evident that the configuration KPCA with two components exhibits the lowest Davies-Bouldin Score (0.00), signifying the formation of highly distinguishable clusters. In contrast, configurations based on LPP also yield a Davies-Bouldin Score of zero, which reinforces the earlier observation. This observation suggests that the Davies-Bouldin Score may not be an appropriate measure when LPP is employed as a dimensionality reduction method.

The Calinski-Harabasz Score measures the ratio of between-cluster variance to within-cluster variance, with higher values indicating better clustering. It quantifies the dispersion between clusters relative to the dispersion within clusters. Looking at Table 4, we find that KPCA with two components also achieves the highest Calinski-Harabasz Score ($4.33 \times 10^9$), demonstrating its effectiveness in generating clusters with minimal within-cluster variance.

In summary, the clustering results show that feature reduction methods like ICA and KPCA tend to yield better clustering performance, with higher Silhouette, lower Davies-Bouldin, and higher Calinski-Harabasz Scores. Conversely, LPP-based configurations perform less optimally, particularly in achieving well-separated clusters as indicated by lower Silhouette and higher Davies-Bouldin Scores.

*4.3.3. Hierarchical clustering with seven segments*

Table 5 presents the results of clustering index analysis for various feature reduction and clustering configurations with 7 segments. The evaluation

Table 4. Results for clustering verification indices - Configuration 5-segments.

| DRT | # of components | Sil. score | CH. score | DB. Score |
|---|---|---|---|---|
| ICA | | 0.76 | 1879.69 | 0.52 |
| KPCA | 2 components | 1.00 | $4.33 \times 10^9$ | 0.00 |
| SVD | | 0.76 | 1555.86 | 0.39 |
| LPP | | 0.72 | 3963.03 | 0.00 |
| ICA | | 0.83 | 114.27 | 0.51 |
| KPCA | 10 components | 0.99 | 1229.01 | 0.01 |
| SVD | | 0.75 | 63.86 | 0.40 |
| LPP | | 0.74 | 1138.88 | 0.00 |
| ICA | | 0.87 | 106.91 | 0.07 |
| KPCA | 20 components | 0.74 | 63.84 | 0.40 |
| SVD | | 0.75 | 63.86 | 0.40 |
| LPP | | 0.76 | 138.20 | 0.00 |
| ICA | | 0.82 | 40.86 | 0.12 |
| KPCA | 40 components | 0.98 | 877.62 | 0.01 |
| SVD | | 0.75 | 63.86 | 0.40 |
| LPP | | 0.76 | 138.20 | 0.00 |
| Original | 80 components | 0.75 | 63.86 | 0.40 |

Table 5. Results for clustering verification indices - Configuration 7-segments.

| DRT | # of components | Sil. score | CH. score | DB. Score |
|---|---|---|---|---|
| ICA | | 0.75 | 1300.62 | 0.54 |
| KPCA | 2 components | 0.99 | 437626082735469.5 | 0.00 |
| SVD | | 0.74 | 1140.63 | 0.42 |
| LPP | | 0.71 | 2864.90 | 0.00 |
| ICA | | 0.81 | 85.52 | 0.40 |
| KPCA | 10 components | 0.99 | 2125.56 | 0.01 |
| SVD | | 0.73 | 153.61 | 0.42 |
| LPP | | 0.72 | 862.50 | 0.00 |
| ICA | | 0.84 | 87.80 | 0.08 |
| KPCA | 20 components | 0.84 | 87.80 | 0.08 |
| SVD | | 0.73 | 153.61 | 0.42 |
| LPP | | 0.73 | 655.09 | 0.00 |
| ICA | | 0.72 | 67.21 | 0.32 |
| KPCA | 40 components | 0.98 | 1215.33 | 0.01 |
| SVD | | 0.73 | 153.61 | 0.42 |
| LPP | | 0.73 | 655.09 | 0.00 |
| Original | 80 components | 0.73 | 153.61 | 0.42 |

metrics include the Silhouette Score, Davies-Bouldin Score, and Calinski-Harabasz Index.

First, we observe that the Silhouette Score, which measures the separation between clusters, varies across configurations. The ICA with twenty components configuration achieves a relatively high Silhouette Score of 0.84, indicating well-defined clusters. Conversely, the KPCA with two components configuration exhibits a Silhouette Score of 0.9999, suggesting almost perfect cluster separation. Second, the Davies-Bouldin Score quantifies the average similarity between each cluster and its most similar cluster, with lower scores indicating better cluster separation. Notably, the KPCA with two components configuration has an extremely low Davies-Bouldin Score of 0.0000, indicating highly distinct clusters. Finally, the Calinski-Harabasz Index measures the compactness and separation of clusters. The ICA with ten components configuration shows a Calinski-Harabasz Index of 0.40, reflecting its effectiveness in forming compact yet well-separated clusters. In contrast, KPCA with two components stands out with a Calinski-Harabasz Index of 0.00, suggesting the formation of highly compact and well-separated clusters.

### 4.3.4. Hierarchical clustering with nine segments

From Table 6, we can analyze the performance of different dimensionality reduction techniques on 9 segments clustering. KPCA with two components stands out with a Silhouette Score of 1.00, indicating well-separated clusters. However, it has an extremely high Calinski-Harabasz Index ($3.0019 \times 10^{27}$) and a low Davies-Bouldin Index of 0.00. In contrast, ICA with twenty components also shows promising results with a Silhouette Score of 0.81, suggesting good cluster quality. It achieves a relatively low Calinski-Harabasz Index (77.94) and a Davies-Bouldin Index of 0.15. On the other hand, LPP-based configurations LPP with 2, 10, 20, and 40 components exhibit Silhouette Scores below 0.71, indicating less well-defined clusters. Furthermore, they all share a Davies-Bouldin Index of 0.00, suggesting that this index may not be suitable for evaluating clustering results when LPP is applied as the dimensionality reduction method.

Overall, KPCA with two components and ICA with 20 components perform well in terms of Silhouette Score, but their suitability may depend on the specific requirements of the clustering task.

### 4.4. Results summary

In this subsection, we summarize the results of applying various dimensionality reduction techniques and clustering methods across different segmentation scenarios. Tables 3–6 provide insights into the performance of these methods.

Table 3 examines the results for 3 segments clustering. KPCA with two components stands out with a nearly perfect Silhouette Score of 0.9992, indicating excellent cluster separation. However, LPP-based configurations exhibit lower Silhouette Scores, suggesting less distinct clusters. Additionally, KPCA with two components achieves the highest Calinski-Harabasz Index, emphasizing its effectiveness in forming compact and separated clusters. Conversely, the Davies-Bouldin Index yields zero for LPP configurations, which may be attributed to extreme dimensionality reduction and well-separated clusters. Table 4 focuses on 5 segments clustering, where ICA with twenty components and KPCA with two components achieve the best Silhouette Scores. KPCA with two components notably attains a perfect score, indicating near-perfect clustering. However, a Davies-Bouldin Score of zero for LPP configurations raises concerns about the suitability of this index when LPP is used as the dimensionality reduction method.

Moving to 7 segments clustering in Table 5, we observe that KPCA with two components leads with the highest Silhouette and lowest Davies-Bouldin Scores, indicating highly distinct clusters. ICA with twenty components performs well, achieving a high Silhouette Score and low Davies-Bouldin Score, while ICA with 10 components exhibits a good balance between separation and compactness. In Table 6, which focuses on 9 segments clustering, KPCA two components again stands out with a perfect Silhouette Score. However, its extremely high Calinski-Harabasz Index suggests extreme compactness, which may not align with all clustering objectives. ICA with twenty components showcases a good Silhouette Score and balanced clustering quality. Fig. 3 provides a comprehensive overview and comparison of various runs with different segmentation strategies and distinct DRTs, all employing the same number of components (2 components). The depicted results reveal noteworthy trends and disparities among the evaluated techniques.

Table 6. Results for clustering verification indices - Configuration 9-segments.

| DRT | # of components | Sil. score | CH. score | DB. Score |
|---|---|---|---|---|
| ICA | | 0.68 | 1427.71 | 0.57 |
| KPCA | 2 components | 1.00 | $3 \times 10^{27}$ | 0.00 |
| SVD | | 0.69 | 1561.52 | 0.54 |
| LPP | | 0.71 | 2185.07 | 0.00 |
| ICA | | 0.80 | 89.45 | 0.46 |
| KPCA | 10 components | 0.98 | 3072.57 | 0.01 |
| SVD | | 0.72 | 443.00 | 0.64 |
| LPP | | 0.70 | 763.75 | 0.00 |
| ICA | | 0.81 | 77.94 | 0.15 |
| KPCA | 20 components | 0.73 | 443.00 | 0.64 |
| SVD | | 0.73 | 443.00 | 0.64 |
| LPP | | 0.71 | 543.37 | 0.00 |
| ICA | | 0.73 | 578.44 | 0.58 |
| KPCA | 40 components | 0.98 | 1245.76 | 0.01 |
| SVD | | 0.72 | 443.00 | 0.64 |
| LPP | | 0.71 | 543.37 | 0.00 |
| Original | 80 components | 0.72 | 443.00 | 0.64 |

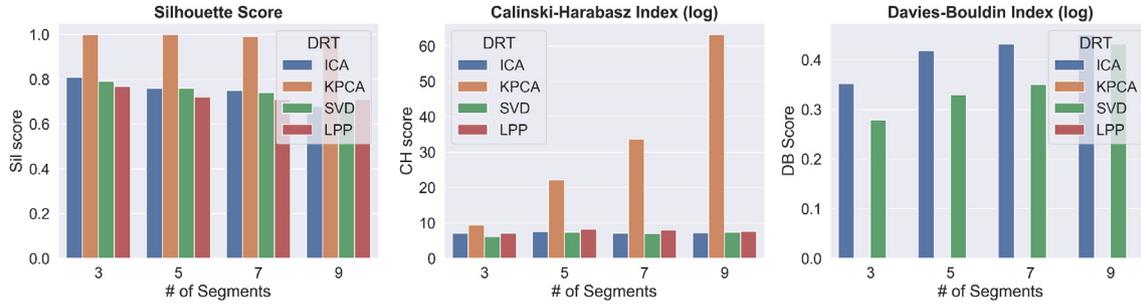

Fig. 3. Summary and comparison of DRT's with various segments and consist of two components.

The analysis shows that KPCA consistently exhibits near-perfect scores across all evaluation metrics, suggesting its effectiveness in generating well-defined clusters. In contrast, LPP demonstrates suboptimal results, particularly in the first two scores, and becomes unmeasurable in the third score. The detailed examination of the 2-component scenario is presented graphically, providing a sample representation of the results. Comprehensive tabular details for all runs are available for reference. Both KPCA and LPP have a zero score in the last measure, known as the DB score. However, the underlying reasons for this outcome differ. In the case of LPP, the measure is not quantifiable, as previously mentioned in the research. On the other hand, KPCA achieves near-perfect performance, leading to its zero score in the DB measure.

In summary, it is observed that KPCA, especially with a low number of features, consistently yields nearly perfect scores across all evaluation indices when combined with AHC. On the contrary, LPP faces challenges in terms of validation, particularly when using the Davies-Bouldin score. This difficulty can be attributed to the unique nature of the reduced data generated by LPP. Additionally, when the original data, without any dimensionality reduction, is utilized, it exhibits relatively poor performance compared to most Dimensionality Reduction Techniques. This highlights the profound impact of the curse of dimensionality and underscores the necessity of employing DRTs prior to applying clustering techniques.

## 5. Conclusion

In the pursuit of enhancing Anti-Money Laundering systems, this study delved into the realm of dimensionality reduction techniques (DRTs) and clustering methodologies. AML, being paramount in safeguarding financial integrity, requires the identification of high-risk groups based on transactional behavior. However, the high dimensionality of behavioral data poses significant challenges to clustering performance. The investigation revolved around the amalgamation of Agglomerative Hierarchical Clustering (AHC) with four prominent DRTs: Independent Component Analysis (ICA), Kernel Principal Component Analysis (KPCA), Singular Value Decomposition (SVD), and Locality Preserving Projections (LPP). The primary objective was to mitigate the adversities posed by high-dimensional AML data and to elevate the precision of clustering-based AML systems.

The study's findings illuminate KPCA as the star performer among DRTs when integrated with AHC. Across various segmentation scenarios, KPCA consistently yields exceptional results, as underscored by three distinct validation indices. Its proficiency in forming well-separated and compact clusters, as evidenced by high Silhouette Scores and low Davies-Bouldin Scores, accentuates its excellence in handling AML data's dimensionality. Conversely, LPP exhibited peculiar behaviors that rendered its validation via the Davies-Bouldin score problematic. The extreme dimensionality reduction imposed by LPP led to zero scores, potentially masking valuable insights into cluster quality. This unique challenge highlights the need for careful consideration when employing LPP as a DRT within AML systems. Additionally, the study revealed the detrimental impact of high dimensionality on clustering performance, exemplified when the original, unreduced data displayed subpar results compared to DRT-enhanced approaches. This underscores the paramount importance of DRTs as a crucial pre-processing step to mitigate the curse of dimensionality and optimize the efficiency of clustering techniques in AML.

In conclusion, this research advocates for the strategic integration of KPCA with AHC as a potent solution for improving the clustering performance of AML systems. It also sheds light on the intricate dynamics of dimensionality reduction techniques within this context, offering valuable insights for the ongoing enhancement of AML strategies. The findings herein pave the way for more effective AML systems that can better identify high-risk behaviors and contribute to the sustained integrity of financial systems worldwide.

Future work will involve exploration of alternative clustering techniques such as K-means and DBSCAN, to validate the robustness of the observed results. Investigating whether these techniques yield similar outcomes or offer distinct insights into clustering behavior can provide a comprehensive understanding of the most suitable clustering approach for AML data.

## Conflict of interest

The authors declare no competing interests.